\documentclass[preprint2]{proto}
\usepackage{times}

\newcommand{\refs}{\par\noindent\hangindent=1pc\hangafter=1}
\begin{document}

\title{\textbf{\LARGE The Dynamical Evolution of the Asteroid Belt}}

\author {\textbf{\large Alessandro Morbidelli}}
\affil{\small\em Dep. Lagrange, CNRS, Observatoire de la Cote d'Azur, Universit\'e de Nice Sophia-Antipolis; Nice, France }
\author {\textbf{\large Kevin J. Walsh}}
\affil{\small\em Southwest Research Institute; Boulder, Co.}
\author {\textbf{\large David P. O'Brien}}
\affil{\small\em Planetary Science Institute; Tucson, Az.}
\author {\textbf{\large David A. Minton}}
\affil{\small\em Purdue University, Department of Earth, Atmospheric, and Planetary Sciences; West Lafayette, IN}
\author {\textbf{\large William F. Bottke}}
\affil{\small\em Southwest Research Institute; Boulder, Co.}

\begin{abstract}
\begin{list}{ } {\rightmargin 1in}
\baselineskip = 11pt
\parindent=1pc
{\small The asteroid belt is the leftover of the original planetesimal population in the inner solar system. However, currently the asteroids have orbits with all possible values of eccentricities and inclinations compatible with long-term dynamical stability, whereas the initial planetesimal orbits should have been quasi-circular and almost co-planar. The total mass now contained in the asteroid population is a small fraction of that existing primordially. Also, asteroids with different {chemical/mineralogical properties are not ranked in an orderly manner with mean heliocentric distance (orbital semi major axis) as one could expect from the existence of a radial gradient of the temperature in the proto-planetary disk,} but they are partially mixed. These properties show that the asteroid belt has been severely sculpted by one or a series of processes during its lifetime. This paper reviews the processes that have been proposed so far, discussing the properties that they explain and the problems that they are confronted with. Emphasis is paid to the interplay between the dynamical and the collisional evolution of the asteroid population, which allows the use of the size distribution or of the crater densities observed in the asteroid belt to constrain the dynamical models.  We divide the asteroid belt evolution into three phases. The first phase started during the lifetime of the gaseous proto-planetary disk, when the giant planets formed and presumably experienced large-scale migrations, and  continued after the removal of the gas, during the build-up of the terrestrial planets. The second phase occurred after the {removal of the gaseous proto-planetary disk} and it became particularly lively for the asteroid belt when the giant planets suddenly changed their orbits, as a result of a mutual dynamical instability and the interaction with the trans-Neptunian planetesimal disk. The third phase covers the aftermath of the giant planet instability, until today. 
 \\~\\~\\~}

\end{list}
\end{abstract}

\section{\textbf{INTRODUCTION}}

The asteroid belt helps us in reconstructing the origin and the evolution of the Solar System, probably better than the planets themselves. This is because the asteroid belt provides several key constraints that can be used effectively to guide the development, the calibration and the validation of evolutionary models. Compared to other small body populations, such as the Kuiper belt or the Oort cloud, the constraints provided by the asteroid belt are probably more stringent, due to the fact that the number and the properties of the asteroids are better known, thanks to ground based observations, space missions and meteorite analysis.  

The structure of this review chapter is therefore as follows. We start by reviewing in section 2 what the most important observational constraints on the asteroid belt are and what they suggest. Then, in section 3, we will review the main models proposed, from the oldest to the most recent ones, and from the earliest to the latest evolutionary phases they address. In Section 4, we will discuss several implications for asteroid science from our current preferred view of the dynamical evolution of the asteroid belt.

\begin{figure*}[t!]
 \epsscale{1.5}
 \plotone{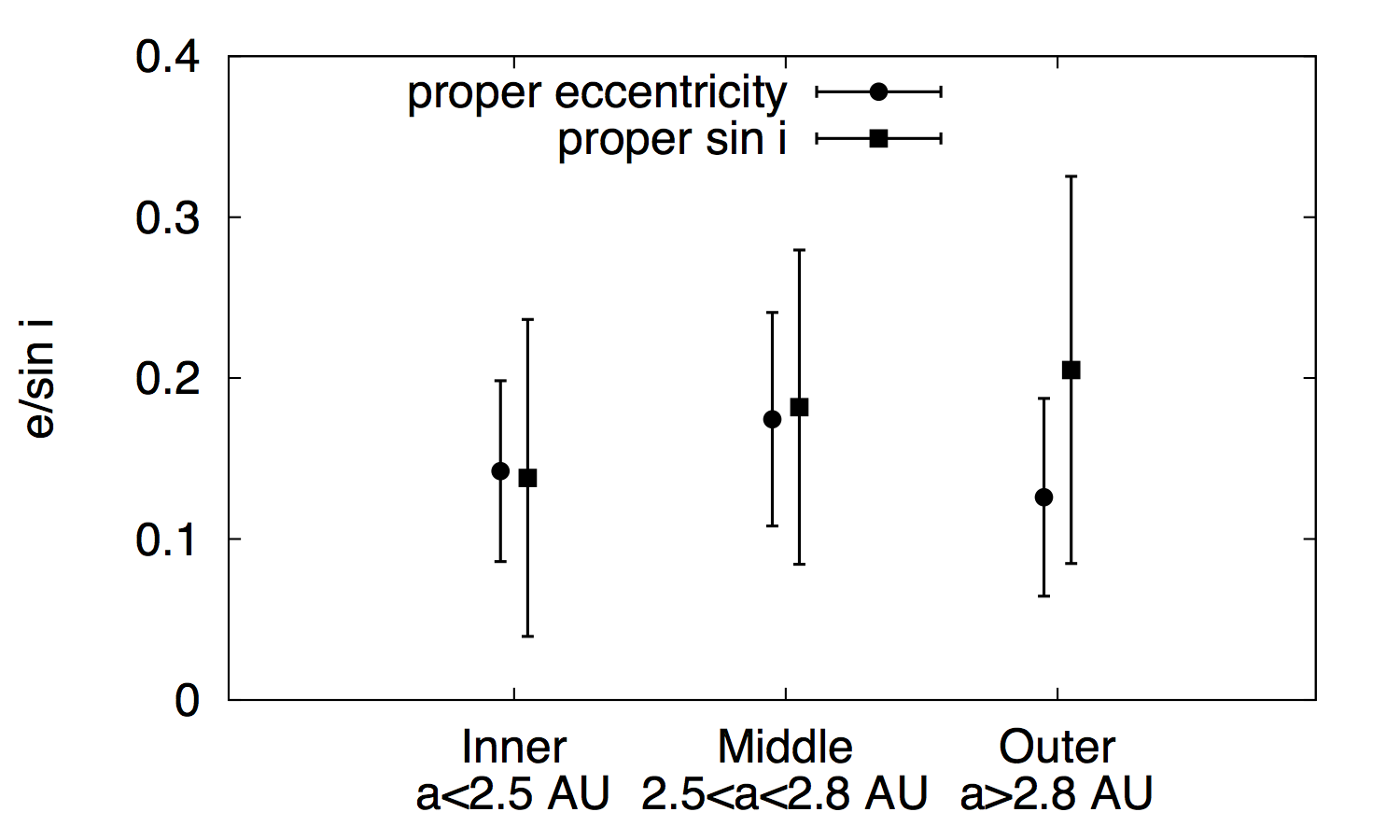}
 \caption{\small The points show mean proper eccentricity (circles) and mean proper inclination (squares) for the $D>100$ km asteroids, divided into three bins of semi major axis. The error bars show the 1-$\sigma$ standard deviation. There is little systematic difference in excitation across the main belt. The slightly increase of inclination from the inner to the outer belt is due to the effect of the $g=g_6$ secular resonance (see sect. 3), which most strongly affects high inclination asteroids in the inner belt.}
\end{figure*}
 
The dynamical evolution of the asteroid belt has already been the object of a review chapter by Petit et al. (2002) in the Asteroid III book. This review has therefore an important overlap with that chapter. Nevertheless, both our observational knowledge of the asteroid belt and our theoretical understanding of Solar System evolution have improved significantly since the early 2000, providing an emerging view of a very dynamic early Solar System, in which various episodes of planet migration played a fundamental role in sculpting the small body reservoirs and displacing planetesimals far from their original birth places. Thus this chapter will present {in greater details models proposed after 2002}, focusing on their implications for asteroid science. Moreover, when reviewing models already presented in Petit et al., we will refer to numerical simulations of these models made after the Petit et al. chapter. 

\section{\textbf{Observational constraints on the primordial evolution of the asteroid belt}}

The observational constraints most useful for reconstructing the formation and evolution of the asteroid belt are those related to large asteroids (larger than $\sim 50$--$100$ km in diameter). In fact, it has been argued that these asteroids are the most likely to be ``pristine'' in the sense that they were not  generated in large numbers in collisional break-up events of larger parent bodies (Bottke et al. 2005a; see chapter by Bottke et al.), nor have they been affected by gas drag and other non-gravitational forces (e.g., the Yarkovsky effect; see chapter by Vokrouhlicky et al.). Moreover, there is an emerging view that the first planetesimals were big, with a preferred diameter in the range mentioned above (Morbidelli et al., 2009; see chapter by Johansen et al.). Thus, throughout this chapter we will limit our discussion to the properties of large asteroids and refer to smaller asteroids only when explicitly mentioned. 

A key major characteristic of the asteroid belt population is the orbital excitation, i.e. the fact that the eccentricities and inclinations of {many} asteroidal orbits are quite large (e.g, Petit et al. 2002). The median proper inclination of $D>100$~km asteroids is 11 degrees and the median proper eccentricity is 0.145. More importantly, the values of eccentricities and inclinations of the largest asteroids are considerably dispersed, with the former ranging between 0 and  0.30, while the latter ranges between 0 and 33 degrees (see Fig. 1).   It has been shown that asteroids of modest inclinations ($i < 20^\circ$) fill the entire orbital space available for long-term dynamical stability, though some stable regions are more densely populated than others (Minton and Malhotra, 2009, 2011). The reader should be aware that, whatever the preferred formation mechanism (see Johansen et al. chapter), planetesimals are expected to have formed on circular and co-planar orbits. Thus, one or more dynamical excitation mechanism(s) within the primordial asteroid belt were needed to stir up eccentricities and inclinations to randomly dispersed values. Asteroid eccentricities and inclinations do not show a strong dependence on semi major axis (Fig. 1)

\begin{figure*}[t!]
 \epsscale{1.5}
 \plotone{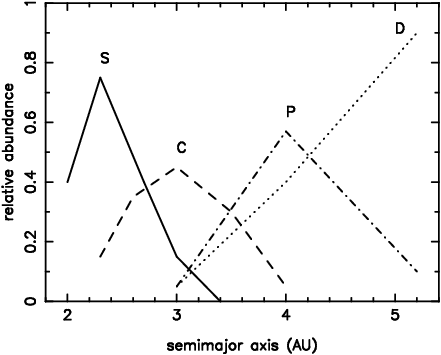}
 \caption{\small The relative distribution of large asteroids ($D>50$~km) of different taxonomic types as originally observed by Gradie and Tedesco (1982). Further works by Moth\'e-Diniz et al. (2003), Carvano et al. (2010) and DeMeo and Carry (2014) demonstrate that the level of mixing increases for smaller asteroid sizes.}  
 \end{figure*}

A second fundamental characteristic of the asteroid belt is the partial mixing of taxonomic classes. Asteroids can be grouped into many taxonomic classes on the basis of their visual and infra-red spectroscopic signatures (Tholen, 1984; Bus and Binzel, 2002; DeMeo et al., 2009). As shown first by Gradie and Tedesco (1982) for the largest asteroids, the inner belt is dominated by S-complex asteroids, many of which are probably related to the meteorites known as ordinary chondrites (Binzel et al. 1996; see also the chapter by Vernazza et al.). The central belt (2.5-3.2 AU) is dominated by C-complex asteroids, probably related to carbonaceous chondrites (Burbine et al., 2002; see also the chapters by DeMeo et al. and Rivkin et al.). The Cybeles asteroids (3.2-3.7 AU), the Hilda asteroids (in the 3/2 mean motion resonance with Jupiter)  and the Jupiter Trojan asteroids (in the 1/1 resonance with Jupiter) are dominated by P-and D-type asteroids (see chapter by Emery et al.).  The C2 ungrouped meteorite ``Tagish Lake'' has been proposed to be a fragment of a D-type asteroid (Hiroi et al., 2001).

This stratification of the main belt makes intuitive sense in terms of a general view that proto-planetary disks should have temperatures decreasing with increasing distance from the central star. In fact, ordinary chondrites are less abundant in {organics} and water than carbonaceous chondrites and therefore are more likely to have formed in a warmer part of the disk. {The small water content in ordinary chondrites, well below the solar proportion, suggests that these bodies accreted closer to the Sun than the snowline. {The fact that some water is nevertheless present is not in contradiction with this statement. A small amount of water could have been accreted by collisions with primitive bodies scattered or drifting into the inner part of the disk}}. At the opposite extreme, the CI meteorites show no chemical fractionation relative to the solar composition,  except {H, C, N, O and all noble gases}, suggesting that they formed in a region of the disk where the temperature was low enough to allow the condensation of most elements. 

As shown in Fig. 2, however, asteroids of different taxonomic types are partially mixed in orbital semi major axis, which smears the trend relating physical properties to heliocentric distance. The mixing of taxonomic type should not be interpreted as the existence of asteroids of intermediate physical properties between those of adjacent types; instead it is due to the coexistence of asteroids of different types with various relative proportions at each value of semi major axis. {Some mixing could come from the fact that the thermal and chemical compositional properties of the disk probably changed over time. However, given that no systematic differences in accretion ages is observed among the main group of chondrites (Villeneuve et al., 2009), it is more likely that some mechanism}, possibly the same that excited the orbital eccentricities and inclinations, also changed somewhat in a random fashion the original semi major axes of the bodies, causing the observed partial mixing.

The asteroid belt contains overall very little mass. From the direct determination of the masses of the largest asteroids and an estimate of the total mass of the ring of bodies which cannot be individually "weighted", based on the collective gravitational perturbations exerted on Mars, Krasinsky et al. (2002), Kuchynka and Folkner (2013) and Somenzi et al. (2010) concluded that the total mass contained in the asteroid belt is $\sim4.5\times10^{-4}$ Earth masses ($M_\oplus$). This value is very low compared to that estimated to have originally existed in the primordial asteroid belt region by interpolating the mass densities required to form the terrestrial planets and the core of Jupiter at both ends of the belt (Weidenschilling, 1977), which is of the order of 1~$M_\oplus$ {(within a factor of a few)}.  Thus, the mass in the asteroid belt has potentially been depleted by three orders of magnitude compared to these expectations. 

We can glean insights into how the primordial belt lost its mass by investigating what we know about its collisional evolution. The collisional history of asteroids is the subject of the chapter by Bottke et al., but we report the highlights here that are needed for this discussion. {In brief, using a number of constraints, Bottke et al. (2005a) concluded that the integrated collisional activity of the asteroid belt is equivalent to the one that would be produced at the current collisional rate over 8-10~Gy.} 

This result has several implications. First, it strongly suggests that the three orders of magnitude mass depletion could not come purely from collisional erosion; such intense comminution would violate numerous constraints. Second, it argues that the mass depletion of the asteroid belt occurred very early. This is because, once the eccentricities and inclinations are excited to values comparable to the current ones, {for a given body every million year spent in an asteroid belt 1,000 times more populated brings a number of collisions equivalent to that suffered in 1 Gy within the current population}. For this reason, the third implication is that the dynamical excitation and the mass depletion event almost certainly coincided. This argues that the real dynamical excitation event was stronger than suggested by the current distribution of  asteroid eccentricities.  One way to reconcile  a massive asteroid belt with this scenario is to assume that  over 99\% of the asteroids had their orbits so excited that they left the asteroid belt forever (hence the mass depletion).  This would make the eccentricities (and to a lesser extent the inclinations) we see today to be those defined by the the lucky survivors, namely the bodies whose orbits were excited the least. 

Using these constraints, we discuss in the next session the various models that have been proposed for the primordial sculpting of the asteroid belt. 

\section{\textbf{Models of primordial evolution of the asteroid belt}}

\subsection{\textbf{Early models}}

The first attempts to explain the primordial dynamical excitation of the asteroid belt were made by Heppenheimer (1980) and Ward (1981) who proposed that secular resonances swept through the asteroid belt region during the dissipation of gas in the proto-planetary disk. Secular resonances occur when the precession rate of the orbit of an asteroid is equal to one of the fundamental frequencies of precession of the orbits of the planets. There are two angles that characterize the orientation of an orbit in space, the longitude of perihelion ($\varpi$) and the longitude of the ascending node ($\Omega$), each of which can precess at different rates depending on the gravitational effects of the other planets and nebular gas (if present). The resonances that occur when the precession rates of the longitudes of perihelion of an asteroid (denoted by $g$) and of a planet are equal to each other excite the asteroid's eccentricity. Similarly, the resonances occurring when the precession rates of the longitudes of node of an asteroid (denoted by $s$) and of a planet are equal to each other excite the asteroid's inclination. In the case of asteroids in the main belt, the planets' precession frequencies that most influence their dynamics are those associated with the orbits of Jupiter and Saturn.  These are called $g_5$ and $g_6$ for the longitude of perihelion precession (the former dominating in the precession of the perihelion of Jupiter, the latter in that of Saturn), and $s_6$ for the longitude of the node precession (both the nodes of Jupiter and Saturn precess at the same rate, if defined relative to the {\it invariable plane}, defined as the plane orthogonal to their total angular momentum vector).

The dissipation of gas from the proto-planetary disk changes the gravitational potentials that the asteroids and planets feel, and hence changes the precession rates of their orbits.  Given that the planets and asteroids are at different locations, they will be affected somewhat differently by this change of gravitational potential and consequently their precession rates will not change proportionally. It is therefore possible that secular resonances sweep through the asteroid belt as the gas dissipates. This means that every asteroid, whatever its location in the belt, first has orbital precession rates slower than the $g_5, g_6$ frequencies of Jupiter and Saturn when there is a lot of gas in the disk, then enters resonance ($g=g_5$ or $g=g_6$) when some appropriate fraction of the gas has been removed, and eventually is no longer in resonance (its orbital precession frequency being faster than those of the giant planets: i.e. $g>g_6$) after all the gas has disappeared. The same occurs for the asteroid's nodal frequency $s$ relative to the planetary frequency $s_6$. This sweeping of perihelion and nodal secular resonances has the potential to excite the orbital eccentricities and inclinations of all asteroids.

This mechanism of asteroid excitation due to disk dissipation has been revisited with numerical simulations in Lemaitre and Dubru (1991), Lecar and Franklin (1997),  Nagasawa et al. (2000,2001,2002), Petit et al. (2002), and finally by O'Brien et al. (2007). Nagasawa et al. (2000) found that of all the scenarios for gas depletion they studied (uniform depletion, inside-out, and outside-in), inside-out depletion of the nebula was most effective at exciting eccentricities and inclinations of asteroids throughout the main belt.  However, they (unrealistically) assumed that the nebula coincided with the ecliptic plane. {Proto-planetary disks can be warped, but they are typically aligned with the orbit of the locally dominant planet (Mouillet et al., 1997). Thus, there is no reason that the gaseous disk in the asteroid belt region was aligned with the current orbital planet of the Earth (which was not formed yet). Almost certainly it was aligned with the orbits of the giant planets. Taking the invariable plane (the plane orthogonal to the total angular momentum of the Solar System) as a proxy of the original orbital plane of Jupiter and Saturn,}   Nagasawa et al. (2001, 2002) found that the excitation of inclinations would be greatly diminished.  Furthermore, since nebular gas in the inside-out depletion scenario would be removed from the asteroid belt region before the resonances swept through it, there would be no gas drag effect to help deplete material from the main belt region.

The work of O'Brien et al. (2007) accounted for the fact that the giant planets should have had orbits significantly less inclined and eccentric than their current values when they were still embedded in the disk of gas, because of the strong damping that gas exerts on planets (Cresswell et al. 2008; Kley and Nelson, 2012). They concluded that secular resonance sweeping is effective at exciting eccentricities and inclinations to their current values only if gas is removed from the inside-out and very slowly, on a timescale of $\sim$20~My. This gas-removal mode is very different from our current understanding of the photo-evaporation process (Alexander et al., 2014), and inconsistent with observations suggesting that disks around solar-type stars have lifetimes of only 1-10 My, with an average of $\sim$3 My (eg. Strom et al. 1993; Zuckerman et al. 1995; Kenyon and Hartmann 1995; Haisch et al. 2001).

\begin{figure*}[t!]
 \epsscale{2.2}
 \plotone{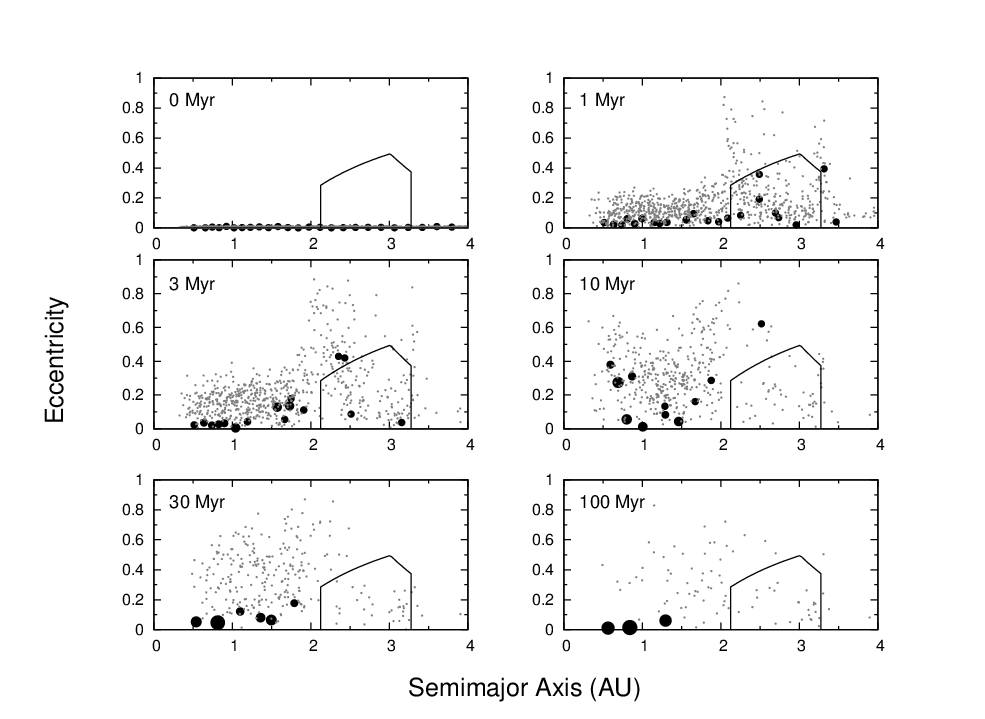}
 \caption{\small Snapshots of the evolution of the solar system and of the asteroid belt in a simulation of Wetherill's model performed in O'Brien et al. (2006) and assuming Jupiter and Saturn on initial quasi-circular orbits. Each panel depicts the eccentricity vs. semi major axis distribution of the particles in the system at different times, labeled on top. Planetesimals are represented with gray dots and planetary embryos by black circles, whose size is proportional to the cubic root of their mass. The solid lines show the approximate boundaries of the current main belt. }  
 \end{figure*}

Earlier studies found that the final eccentricities of the asteroids are quite randomized because two perihelion secular resonances sweep the entire asteroid belt in sequence--first the resonance $g=g_5$, then the resonance $g=g_6$. The first resonance excites the eccentricities of the asteroids from zero to approximately the same value, but the second resonance, sweeping an already excited belt, can increase or decrease the eccentricity depending on the position of the perihelion of each asteroid at the time of the encounter with the resonance (Ward et al., 1976; Minton and Malhotra, 2011).  O'Brien et al. (2007) found that when Jupiter and Saturn were on orbits initially closer together, as predicted by the Nice Model (e.g., Tsiganis et al. 2005), the resonance with frequency $g_6$ would only sweep part of the outer belt, leading to less randomization of eccentricities in the inner belt.  In all studies in which the mid-plane of the proto-planetary disk of gas coincides with the invariable plane of the solar system find that the final inclinations tend to have comparable values. This is because there is only one dominant frequency ($s_6$) in the precession of the nodes of Jupiter and Saturn and hence there is only one nodal secular resonance and no randomization of the final inclinations of the asteroids. Clearly, this is in contrast with the observations. For all these problems, the model of secular resonance sweeping during gas removal {is no longer considered to be able to explain, alone, the excitation and depletion of the primordial asteroid belt}. 

\begin{figure*}[t!]
 \epsscale{2.}
 \plotone{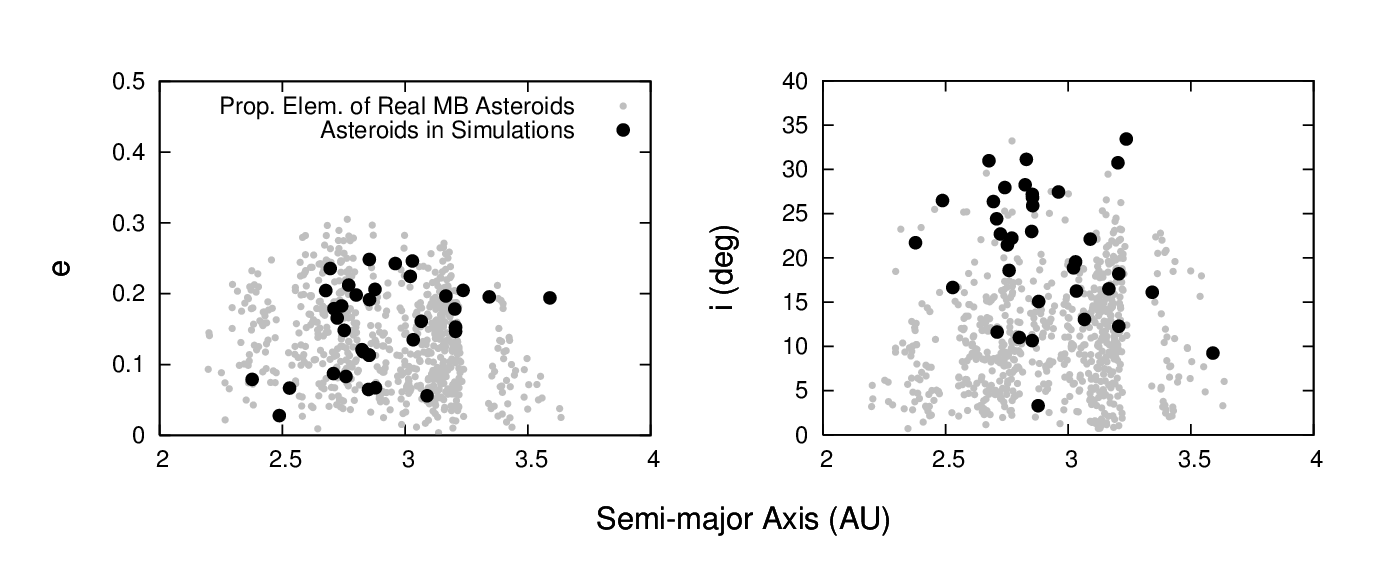}
 \caption{\small The final eccentricities and inclinations of asteroids in Wetherill's (1992) model (black dots), according to the simulations presented in O'Brien et al. (2007). For comparison, the observed distribution of large asteroids is depicted with gray dots. }  
 \end{figure*}

An alternative model for the dynamical excitation of the asteroid belt was proposed by Ip (1987). In this model, putative planetary embryos are scattered out of the Jupiter region and cross the asteroid belt for some timescale before being ultimately dynamically ejected from the Solar System. If the embryos are massive enough, their repeated crossing of the asteroid belt can excite and randomize the eccentricities and inclinations of the asteroids, through close encounters and secular effects. That scenario has been revisited by Petit et al. (1999), who found that, whatever the mass of the putative embryos, the resulting excitation in the asteroid belt ought to be very unbalanced. Excitation would be  much stronger in the outer belt than in the inner belt (because the embryos come from Jupiter's region) and it would be much stronger in eccentricity than in inclination. By contrast, the main asteroid belt shows no such trend (see Fig. 1).  So, again, this model has since been abandoned. {If massive embryos have been scattered from Jupiter's zone, they must have crossed the asteroid belt very briefly so that their limited effects could be completely overprinted by other processes, such as those discussed below.}

\subsection{\textbf{Wetherill's model}}

The first comprehensive model of asteroid belt sculpting, which linked the evolution of the asteroid belt with the process of terrestrial planet formation, was that proposed by Wetherill (1992) and later simulated in a number of subsequent papers (e.g., Chambers and Wetherill, 1998; Petit et al., 2001, 2002; O'Brien et al., 2006, 2007). In this model, at the time gas was removed from the system, the proto-planetary disk interior to Jupiter consisted of a bi-modal population of planetesimals and planetary embryos, the latter with masses comparable to those of the Moon or Mars. Numerical simulations show that, under the effect of the mutual perturbations among the embryos and the resonant perturbations from Jupiter, embryos are generally cleared from the asteroid belt region, whereas embryos collide with each other and build terrestrial planets inside of 2 AU. While they are still crossing the asteroid belt, the embryos also excite and eject most of the original resident planetesimals. Only a minority of the planetesimals (and often no embryos) remain in the belt at the end of the terrestrial planets formation process, which explains the mass depletion of the current asteroid population. The eccentricities and inclinations of the surviving asteroids are excited and randomized, and the remaining asteroids have generally been scattered somewhat relative to their original semi major axes. A series of simulation snapshots demonstrating this process is shown in Figure 3.

Whereas earlier simulations assumed that Jupiter and Saturn were originally on their current orbits, O'Brien et al. (2006, 2007) performed simulations with Jupiter and Saturn on the low-inclination, nearly circular orbits predicted in the Nice Model.  The resulting asteroids from a set of simulations with these initial conditions are shown in Figure 4. Overall, the range of values compare well with those observed for the real asteroids, although the final inclination distribution is skewed towards large inclinations.  The reason for this is that it is easier to excite a low-inclination asteroid to large eccentricity and remove it from the belt than it is for a high-inclination asteroid, because the encounter velocities with the embryos are slower and more effective in deflecting the low-inclination asteroid's orbit.  Also, with the giant planets on nearly circular orbits, it takes longer to clear embryos from the asteroid belt, allowing more time to excite asteroids to large inclinations.

As noted earlier, the surviving asteroids have their orbital semi major axes displaced from their original values, as a result of the embryos' gravitational scattering.  O'Brien et al. (2007) found that the typical change in semi major axis is of the order of 0.5 AU (comparable to earlier simulations), with a tail extending to 1 -- 2 AU, and the semi major axis can be either decreased or increased. This process can explain the partial mixing of taxonomic types. As shown in Fig.~2 the distribution of the S-type and C-type asteroids has a Gaussian-like shape, with a characteristic width of $\sim$0.5 AU. Thus, if one postulates that all S-type originated from the vicinity of 2 AU and all C-type originated in the vicinity of 3 AU, Wetherill's model explains the current distribution.

\subsection{\textbf{The Grand Tack model}}

A more recent, alternative model to Wetherill's is the so-called Grand Tack scenario, proposed in Walsh et al. (2011). Initially the Grand Tack scenario had not been developed to explain the asteroid belt, but to answer two questions left open by Wetherill's model: why is Mars so small relative to the Earth? Why is Jupiter so far from the Sun despite planets having a tendency to migrate inwards in proto-planetary disks? Nevertheless, this scenario has profound implications for the asteroid belt, as we discuss below. 

The Grand Tack scenario is built on results from hydrodynamics simulations finding that Jupiter migrates towards the Sun if it is alone in the gas-disk, while it migrates outward if paired with Saturn (Masset and Snellgrove, 2001; Morbidelli and Crida, 2007; Pierens and Nelson, 2008; Pierens and Raymond, 2011; D'angelo and Marzari, 2012). Thus, the scenario postulates that Jupiter formed first. As long as the planet was basically alone, Saturn being too small to influence its dynamics, Jupiter migrated inwards from its initial position (poorly constrained but estimated at $\sim$3.5 AU) down to 1.5 AU. Then, when Saturn reached a mass close to its current one and an orbit close to that of Jupiter, Jupiter reversed migration direction (aka it "tacked", hence the name of the model) and the pair of planets started to move outwards. This outward migration continued until the final removal of gas in the disk, which the model assumes happened when Jupiter reached a distance of $\sim$5.5 AU. {The migration of the cores of giant planets is still not fully understood (see Kley and Nelson, 2012 for a review). Thus, the Grand Tack model comes in two flavors. In one, Saturn, while growing, migrates inwards with Jupiter. In another, Saturn is stranded at a no-migration orbital radius until its mass exceeds 50~$M_\oplus$ (Bitsch et al., 2014); then it starts migrating inwards and it catches Jupiter in resonance because it migrates faster}. Both versions exist with and without Uranus and Neptune. All these variants are described in Walsh et al. (2011); the results are very similar in all {these} cases, which shows the robustness of the model, {at least withing the range of tested possibilities}. The scheme presented in Fig. 5. has been developed in the framework of the first ``flavor''.

\begin{figure*}[t!]
 \epsscale{1.5}
 \plotone{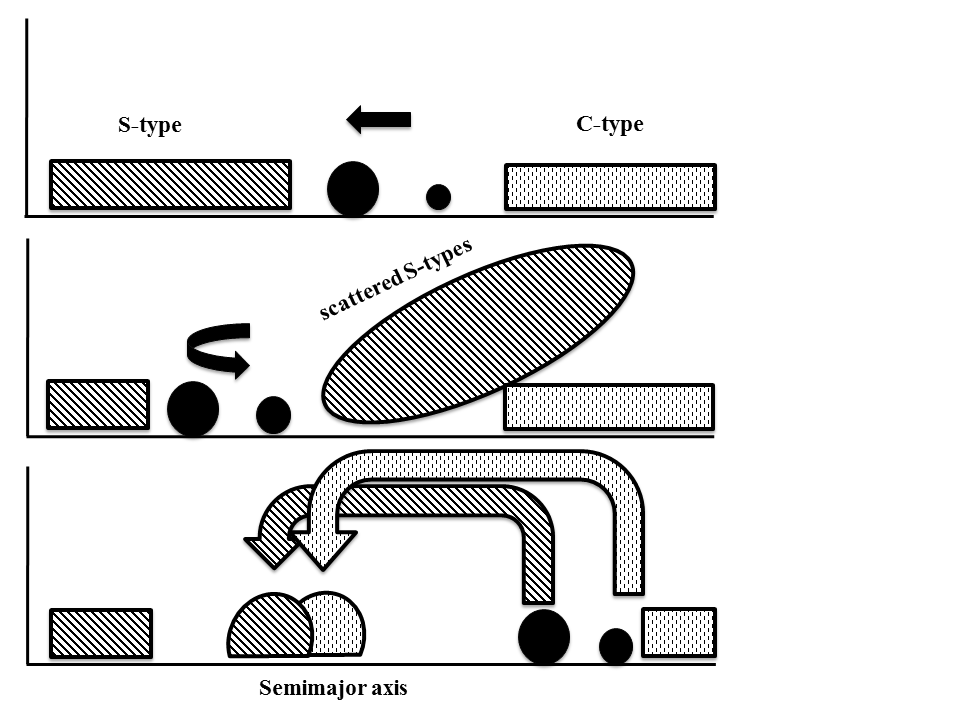}
 \caption{\small A scheme showing the Grand Tack evolution of Jupiter and Saturn and its effects on the asteroid belt. The three panels show three evolutionary states, in temporal sequence. First the planet migrate inwards then, when Saturn reaches its current mass, they move outwards. The dashed and dotted areas schematize the (a,e) distributions of S-type and C-type asteroids respectively. The dashed and dotted arrows in the lower panel illustrate the injection of scattered S-type and C-type asteroids into the asteroid belt during the final phase of outward migration of the planets.}  
 \end{figure*}

\begin{figure*}[t!]
 \epsscale{1.8}
 \plotone{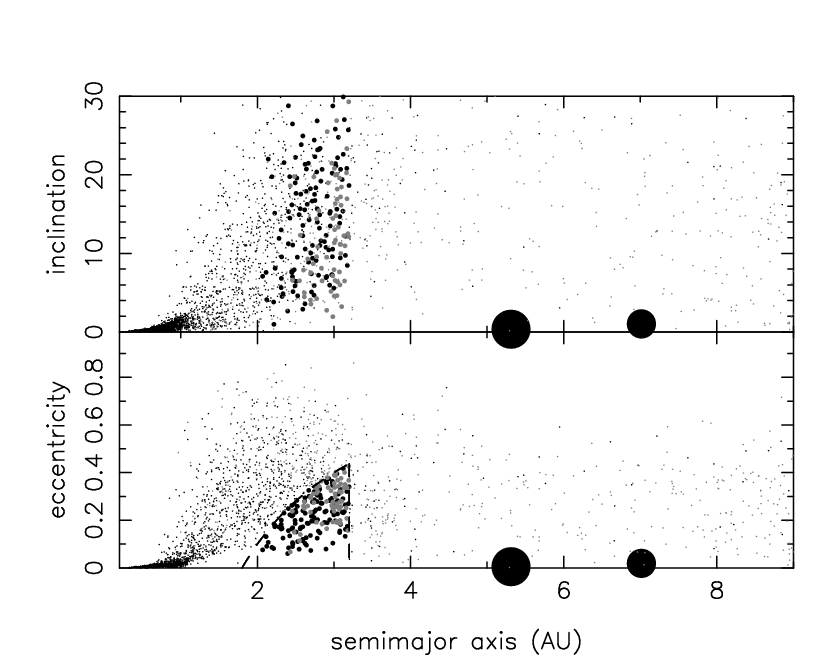}
 \caption{\small Final semi major axis, eccentricity and inclination distribution of bodies surviving the inward and outward migration of Jupiter and Saturn. The black particles were originally placed inside of the initial orbit of Jupiter and the gray particles outside of the initial orbit of Saturn. The particles finally trapped in the asteroid belt are depicted with larger symbols than the others. The dashed curve in the lower panel shows the approximate boundaries of the asteroid belt inward of the 2/1 resonance with Jupiter. This final distribution was achieved in the simulations of Walsh et al. (2011) accounting only for Jupiter and Saturn (i.e., not including Uranus and Neptune) moving together in the 2/3 resonance, as sketched in Fig. 5.}  
 \end{figure*}

Assuming that Jupiter formed at the snowline (a usual assumption to justify the large mass of its core and its fast formation), the planetesimals that formed inside its initial orbit should have been mostly dry. It is therefore reasonable to associate these planetesimals (whose distribution is sketched as a dashed area in Fig. 5) with the S-type asteroids and other even dryer bodies (enstatite-type, Earth precursors etc.). During its inward migration, Jupiter penetrates into the disk of these planetesimals. In doing so, most planetesimals (and planetary embryos) are captured in mean motion resonances with Jupiter and are pushed inwards, increasing the mass density of the inner part of the disk. However, some 10\% of the planetesimals are kicked outwards by an encounter with Jupiter, reaching orbits located beyond Saturn, which collectively have an orbital (a,e) distribution that is typical of a scattered disk (i.e. with mean eccentricity increasing with semi major axis). In semi major axis range, this scattered disk overlaps with the inner part of the disk of primitive bodies (whose distribution is sketched as a dotted area in Fig. 5), which are initially on circular orbits beyond the orbit of Saturn. These bodies, being formed beyond the snowline, should be rich in water ice and other volatile elements, and therefore it is again reasonable to associate them with C-type asteroids. 

\begin{figure*}[t!]
 \epsscale{1.5}
 \plotone{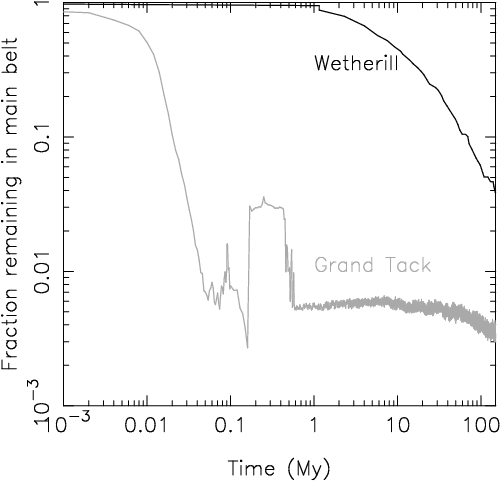}
 \caption{\small The depletion of the asteroid belt in Wetherill's model (black) and Grand Tack model (gray). In the Grand Tack model notice the bump between 0.1 and 0.6~My due to the implantation of primitive objects into the main belt. Overall, the depletion of the asteroid belt is faster and stronger in the Grand Tack model.}  
 \end{figure*}

After reaching $\sim$1.5 AU (this value is constrained by the requirement to form a small Mars and a big Earth; Walsh et al., 2011; Jacobson et al., 2014; Jacobson and Morbidelli, 2014), Jupiter reverses its migration direction and begins its outward migration phase, during which the giant planets encounter the scattered S-type disk, and then also the primitive C-type disk. Some of the bodies in both populations are scattered inwards, reach the asteroid belt region and are implanted there as Jupiter moves out of it. 

The final orbits of the planetesimals, at the end of the outward migration phase, are shown in Fig. 6. A larger dot size is used to highlight the planetesimals trapped in the asteroid belt region and distinguish them from those in the inner solar system or at too large eccentricity to be in the asteroid belt. Notice that the asteroid belt is not empty, although it has been strongly depleted (by a factor of several hundred relative to its initial population). This result is not trivial. One could have expected that Jupiter migrating through the asteroid belt twice (first inwards then outwards) would have completely removed the asteroid population, invalidating the Grand Tack scenario. The eccentricities and the inclinations of the particles in the asteroid belt are excited and randomized. The S-type particles (black) are found predominantly in the inner part of the belt and the C-type particles (gray) in the outer part, but there is a wide overlapping region where both are present. {This is qualitatively consistent with what is observed}.

As discussed above, the Grand Tack scenario solves open problems in Wetherill's model. The small mass of Mars is explained as a result of the disk of the remaining solid material being truncated at $\sim$1 AU (Hansen, 2009; Walsh et al., 2011). Fischer and Ciesla (2014) reported that they could obtain a small-mass Mars in a few percent of simulations conducted in the framework of Wetherill's model. However, the rest of the planetary system in these simulations does not resemble the real terrestrial planet system (Jacobson and Walsh, 2015). For instance another massive planet is formed in Mars-region or beyond. The outward migration of Jupiter explains why the giant planets in our solar system are so far from the Sun, whereas most giant planets found so far around other stars are located at 1-2 AU.  For all these reasons, one can consider the Grand Tack model more as an improvement of Wetherill's model than an alternative, because it is built in the same spirit of linking the asteroid belt sculpting to the evolution of the rest of the solar system (terrestrial planet formation, giant planet migration -- the latter being still unknown at Wetherill's time). 

It is nevertheless interesting to compare the Grand Tack model and Wetherill's model on the basis of the final asteroid belts that they produce. Comparing Fig. 6 with Fig. 4, it is apparent that the Grand Tack model provides a better inclination distribution, more uniform that Wetherill's, but it produces a worse eccentricity distribution, which is now more skewed towards the upper eccentricity boundary of the asteroid belt.

As we will see in Sect.~\ref{Nice}, however, the eccentricity distribution can be remodeled somewhat during a later evolutionary phase of the solar system. This is also partially true also for the inclination distribution. So, for what concerns the eccentricity and inclination distributions one might declare a tie in the competition between the two models. 

The Grand Tack model makes it conceptually easier to understand the significant differences between S-type and C-type asteroids and their respective presumed daughter populations: the ordinary and carbonaceous chondrites. In fact, in the Grand Tack model these two populations are sourced from clearly distinct reservoirs on either sides of the snowline. Instead, in Wetherill's model these bodies would have formed just at the two ends of the asteroid belt, so less than 1 AU apart. Despite such a vast difference in predicted formation locations for these two populations the debate is open. Some authors (e.g. Alexander et al., 2012) think that bodies formed in the giant planet region would be much more similar to comets than to asteroids, others (Gounelle et al., 2008) argue that there is a continuum between C-type asteroids and comets and a clear cleavage of physical properties between ordinary and carbonaceous chondrites. We review the available cosmochemical constraints and their uncertain compatibility with the model in sect.~\ref{cosmo}.

A clear distinction between the Grand Tack model and Wetherill's model is that the former provides a faster and more drastic depletion of the asteroid belt. This point is illustrated in Fig. 7, showing the fraction of the initial asteroid population that is in the main belt region at any time.

The Grand Tack scenario depletes the asteroid belt down to 0.3\%, and does so basically in 0.1~My. Assuming that the final asteroid belt consisted of one current asteroid belt mass in S-type asteroids and three current asteroid belt masses in C-type asteroids (the reason for 4x more total mass in the asteroid belt will be clarified in Sect.~\ref{afterNice}), this implies that the asteroid belt at t=0 should have contained 0.6 Earth masses in planetesimals (the rest in embryos). Also, a calculation of the collision probability of the asteroids as a function of time (both among each other and with the planetesimals outside of the asteroid belt) shows that the integrated collisional activity suffered by the surviving asteroids during the first 200~My would not exceed the equivalent of 4~Gy in the current population. Thus, assuming that the exceeding factor of 4 in the asteroid population is lost within the next 500~My (see Sects.~\ref{Nice} and~\ref{afterNice}), the integrated collisional activity of asteroids throughout the entire solar system age would probably remain within the 10~Gy constraint described in Section 2. In contrast, Wetherill's model depletes the asteroid belt on a timescale of 100~My. Also, about 2-3~\% of the initial population remains in the belt at the end. Thus, to be consistent with constraints on the current population and its integrated collisional activity, the initial mass in planetesimals in the asteroid belt region should have been no larger than 200 times the current asteroid belt mass, or less than one Mars mass (Bottke et al., 2005b). 

\subsection{\textbf{Are cosmochemical constraints consistent with the Grand Tack model?}}
\label{cosmo}

The Grand Tack model predicts that C-type asteroids have been implanted into the asteroid belt from the giant planets region. Is this supported or refuted by cosmochemical evidence? 

Although there is a spread in values, the D/H ratios of carbonaceous chondrites {(with the exception of CR chondrites)} are a good match to Earth's water (Alexander et al., 2012). Oort cloud comets are usually considered to have formed in the giant planet region {(e.g. Dones et al., 2004)}. The D/H ratio was measured {for the water from} seven Oort cloud comets {see Bockelee-Morvan et al., 2012 and references in that paper)}. All but one (comet 153P/Ikeya-Zhang; {Biver et al., 2006}) have water D/H ratios of about twice higher than chondritic. {This prompted Yang et al. (2013) to develop a model where the D/H ratio of ice in the giant planet region is high.} However, Brasser et al. (2007) showed that comet-size bodies could not be scattered from the giant planet region into the Oort cloud in the presence of gas drag (i.e., when the giant planets formed), and Brasser and Morbidelli (2013) demonstrated that the Oort cloud population is consistent with an origin from the primordial trans-Neptunian disk at a later time. {The recent measurement (Altwegg et al., 2014) of a high D/H ratio for the ice of comet 67P/Tchourioumov-Guerassimenko, which comes from the Kuiper belt, supports this conclusion by showing that there is no systematic difference between Oort cloud comets and Kuiper belt comets.} So, care should be taken in using Oort cloud comets as indicators of the D/H ratio in the giant planet region. Conflicting indications on the local D/H ratio come from the analysis of Saturn's moons. Enceladus'  D/H ratio is  roughly twice Earth's (Waite et al. 2009), but Titan's D/H ratio is Earth-like (Coustenis et al. 2008; Abbas et al. 2010; Nixon et al. 2012).

Alexander et al. also noticed a correlation between D/H and C/H in meteorites and interpreted it as evidence for  an isotopic exchange between pristine ice and organic matter within the parent bodies of carbonaceous chondrites. From this consideration, they argued that the original water reservoir of carbonaceous asteroids had a D/H ratio lower  than Titan, Enceladus or any comet, again making asteroids distinct from bodies formed in the giant planet region and beyond. However, a reservoir of pristine ice has never been observed; the fact that Earth's water and other volatiles are in chondritic proportion {(Marty, 2012; however see Halliday, 2013)} means that carbonaceous chondrites --wherever they formed-- reached their current D/H ratios very quickly, before delivering volatiles to Earth. {Possibly, also the D/H ratio measured for comets and satellites might have been the result of a similar rapid exchange between a pristine ice and the organic matter.}

Another isotopic constraint comes from the Nitrogen isotope ratio. Comets seem to have a rather uniform $^{15}$N/$^{14}$N (Rousselot et al., 2014). Even the comets with a chondritic D/H ratio (e.g. Hartley 2; Hartogh et al. 2011)  have a non-chondritic $^{15}$N/$^{14}$N  ratio (Meech et al., 2011). {The $^{15}$N/$^{14}$N ratio, however, is only measured in HCN or NH$_2$, never in molecular Nitrogen N$_2$.} Titan has a cometary $^{15}$N/$^{14}$N as well ({in this case measured in N$_2$; Niemann et al., 2010)}. Here, again, a few caveats are in order. First, it is difficult to relate the composition of a satellite, born from a circum-planetary disk with its own thermal and chemical evolution, to the composition of bodies born at the same solar distance but on heliocentric orbits. Second, it is unclear whether any comets for which isotope ratios have been measured originate from the giant planet region, as opposed to the trans-Neptunian disk (Brasser and Morbidelli, 2013).  We remind that hot-spots with large $^{15}$N/$^{14}$N  ratios are also found in {primitive meteorites (Busemann et al., 2006)}.

Arguments in favor of an isotopic similarity between carbonaceous chondrites and comets come from the analysis of micro-meteorties. Most of micro-meteorites (particles of about $\sim 100\mu$m collected in the Antarctic ice) have chondritic isotopic ratios {for H, C, N, O} (with the exceptions of the ultra-carbonaceous ones, which {have a large D/H ratio, comparable to that in the organics of some chondrites, but which} constitute only a very minority of micrometeorite population - Duprat et al., 2010). Yet, {according to the best available model of the dust distribution in the inner solar system (Nesvorny et al., 2010), which is compelling given that it fits the zodiacal light observations almost perfectly, 
most of the dust accreted by the Earth should be cometary, even when the entry velocity bias is taken into account.}
Similarly, from orbital considerations, Gounelle et al. (2006) concluded that the CI meteorite Orgueil is a piece of a comet. {Also compelling for a continuum between chondrites and comets comes from the examination of comet Wild 2 particles brought to Earth by the Stardust Mission - e.g., Zolensky et al. 2006, 2008.; Ishii et al. 2008;  Nakamura et al., 2008}

These considerations suggest that if one looks at their rocky components, comets and carbonaceous asteroids are very similar from the composition and isotopic point of view, if not indistinguishable. 

Finally, it has been argued that, if the parent bodies of carbonaceous chondrites had accreted among the giant planets they would have contained  $\sim 50$\% water by mass. Instead,  the limited hydrous alteration in carbonaceous meteorites suggests that {only about} $10$\% of the original mass was in water (Krot, 2014; {however see Alexander et al., 2010}). However, a body's original water content cannot easily be estimated from its aqueous alteration. {Even if alteration is complete, there is a finite amount of water that the clays can hold in their structures. Thus, the} carbonaceous parent bodies may have been more water-rich than their alteration seems to imply. In fact, the discoveries of main belt comets {releasing dust at each perihelion passage} (Hsieh and Jewitt, 2006), of water ice on asteroids Themis (Campins et al., 2010; Rivkin and Emery, 2010) and Cybele (Hargrove et al., 2012), and of vapor plumes on Ceres (Kuppers et al., 2014), show that C-type asteroids are more rich in water than their meteorite counterpart seems to suggest, supporting the idea that they might have formed near or beyond the snowline. Meteorites may simply represent rocky fragments of bodies that were far wetter/icier.

Clearly, the debate on whether carbonaceous asteroids really come from the giant planet region as predicted by the Grand Tack model is  wide open. More data are needed from a broader population of comets. The investigation of main belt comets, both remote and in-situ, and the Dawn mission at Ceres will be key to elucidate the real ice-content of carbonaceous asteroids and their relationship with classic "comets".

\subsection{\textbf{The Nice model: a second phase of excitation and depletion for the asteroid belt}}
\label{Nice}

Fig. 7. seems to suggest that after $\sim$100 My the asteroid belt had basically reached its final state. However, at this time the orbits of the giant planets were probably still not the current ones. In fact, the giant planets are expected to have emerged from the gas-disk phase in a compact and resonant configuration as a result of their migration in the gas-dominated disk. This is true not only in the Grand Tack model, but in any model where Jupiter is refrained from migrating inside $\sim$5 AU by whatever mechanism (Morbidelli et al., 2007). 

The transition of the giant planets from this early configuration to the current configuration is expected to have happened via an orbital instability, driven by the interaction of the planets with a massive disk of planetesimals {located from a few AUs beyond the original orbit of Neptune to about 30~AU}  (Morbidelli et al., 2007; Batygin and Brown, 2010; Levison et al., 2011; Nesvorny, 2011; Batygin et al., 2012; Nesvorny and Morbidelli, 2012; {see also a precursor work by Thommes et al., 1999}). In essence, the planetesimals disturbed the orbits of the giant planets and, as soon as two planets fell off resonance, the entire system became unstable.   In the simulations, the instability can occur early (e.g. Morbidelli et al., 2007) or late (Levison et al., 2011) depending on the location of the inner edge of the trans-Neptunian disk. 

\begin{figure*}[t!]
 \epsscale{2.2}
 \plottwo{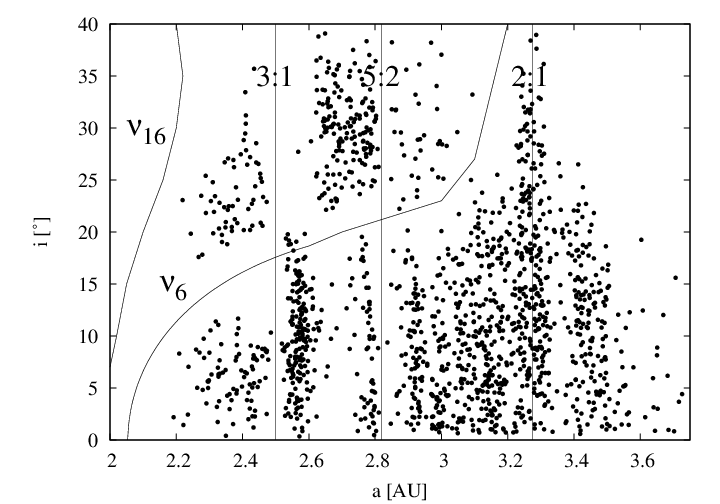}{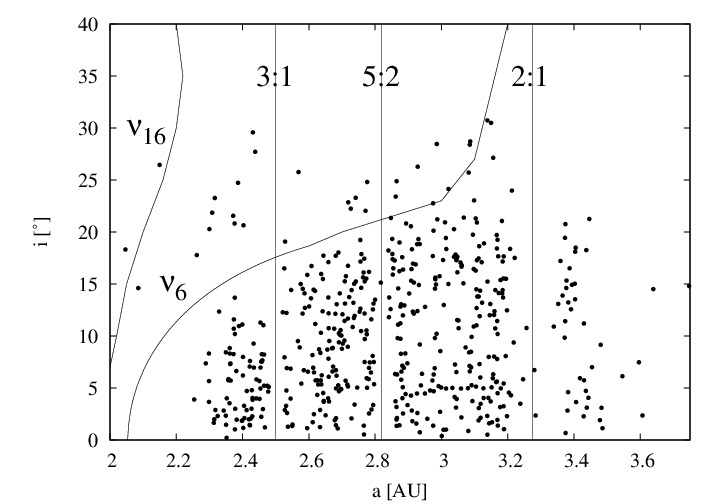}
 \caption{\small A comparison between the final (a,i) distribution of asteroids if Jupiter and Saturn migrate slowly away from each other (left) or jump due to them scattering an ice giant planet (right). In the first case the orbital distribution is inconsistent with that observed, while in the second case it is. From Morbidelli et al. (2010).}  
 \end{figure*}

Constraints suggest that in the real Solar System the instability occurred relatively late, probably around 4.1 Gy ago (namely 450~My after gas removal).  These constraints come primarily from the Moon. {Dating lunar impact basins is difficult, because it is not clear which samples are related to which basin (e.g., Norman and Nemchin, 2014). Nevertheless, it is clear that several impact basins, probably a dozen,} formed in the 4.1-3.8~Gy period (see Fassett and Minton, 2013, for a review). Numerical tests demonstrate that these late basins {(even just Imbrium and Orientale, whose young ages are undisputed)} are unlikely to have been produced by a declining population of planetesimals, left-over from the terrestrial planet accretion process, because of their short dynamical and collisional lifetimes (Bottke et al. 2007). There is also a surge in lunar rock impact ages $\sim$4 Gy ago, which contrasts with a paucity of impact ages between 4.4 and 4.2 Gy (Cohen et al., 2005). This is difficult to explain if the bombardment had been caused by a population of left-over planetesimals slowly declining over time. The situation is very similar for the bombardment of asteroids, with meteorites showing a surge in impact ages 4.1 Gy ago and a paucity of ages between 4.2-4.4 Gy (Marchi et al., 2013). Meteorites also show many impact ages near 4.5 Gy ago, demonstrating that the apparent lack of events in the 4.2-4.4 Gy interval is not due to clock resetting processes. {All these constraints strongly suggest the appearance of a new generation of projectiles in the inner solar system about 4.1 Gy ago, which argues that either a very big asteroid broke up at that time (Cuk, 2012; but such a break-up is very unlikely from collision probability arguments and we don't see any remnant asteroid family supporting this hypothesis), or that the dynamical instability of the giant planets occurred at that time,} partially destabilizing small body reservoirs that had remained stable until then.  

Other constraints pointing to the late instability of the giant planets come from the outer Solar System. If the planets had become unstable at the disappearance of the gas in the disk, presumably the Sun would still have been in a stellar cluster and {consequently the Oort cloud would have formed more tightly bound to the Sun than it is thought to be from the orbital distribution of long period comets (Brasser et al., 2008, 2012)}. Also, the impact basins on Iapetus (a satellite of Saturn) have topographies that have relaxed by 25\% or less, which argues that they formed in a very viscous lithosphere; according to models of the thermal evolution of the satellite, these basins can not have formed earlier than 200~My after the beginning of the Solar System (Robuchon et al., 2011). 

For all these reasons, it is appropriate to discuss the consequences of the giant planet instability on the asteroid belt, after the events described by the Grand Tack or Wetherill's models. In fact, it is important to realize that the model of giant planets' instability (often called the Nice model) is not alternative to the models described before on the early evolution of the asteroid belt; instead it is a model of the subsequent evolution.

The phase of giant planet instability is very chaotic and therefore a variety of orbital evolutions are possible. Nevertheless, the planetary evolutions can be grouped in two categories. In the first category, Saturn, Uranus and Neptune have close encounters with each other, but none of them have encounters with Jupiter. Saturn typically scatters either Uranus or Neptune outwards and thus it recoils inwards. As a result, Uranus and Neptune acquire large eccentricity orbits that cross the trans-Neptunian disk. The dispersal of the planetesimal disk damps the eccentricities of the planets by dynamical friction and drives the planets' divergent migration in semi major axis (Tsiganis et al., 2005). Thus, the planets achieve stable orbits that are well-separated from each other. The orbital separation between Jupiter and Saturn first decreases, when Saturn recoils, and then increases due to planetesimal-driven migration. The timescale for the latter is typically $\sim$~10My. The slow separation between the two major planets of the Solar System drives a slow migration of secular resonances across the asteroid belt (Minton and Malhotra, 2009, 2011) and the terrestrial planet region (Brasser et al., 2009). The problem is that the resulting orbital distribution in the asteroid belt is very uneven, as shown in Fig. 8a, with most asteroids surviving at large inclination (Morbidelli et al., 2010) and the orbits of the terrestrial planets become too excited (Brasser et al., 2009; Agnor and Lin, 2012). 

In the second category of evolutions, Saturn scatters an ice giant planet (Uranus, Neptune or a putative fifth planet) inwards, thus recoiling outwards, and then Jupiter scatters the ice giant outwards, thus recoiling inwards. The interaction with the planetesimals eventually damp and stabilize the orbits of the planets. In this evolution, dubbed 'jumping-Jupiter', the orbital separation between Jupiter and Saturn initially jumps, when Saturn recoils outwards and Jupiter inwards; then there is a final smooth phase of separation, due to planetesimal driven migration. In the jump, the secular resonances can jump across the asteroid belt (Morbidelli et al., 2010) and across the terrestrial planet region (Brasser et al., 2009), without disrupting their orbital structure (see Fig. 8b).

The jumping-Jupiter evolution also explains the capture of the irregular satellites of Jupiter {with an orbital distribution similar to those of the irregular satellites} of the other giant planets (Nesvorny et al., 2007, 2014). It can also explain the capture of Jupiter's Trojans in uneven proportions around the L4 and L5 Lagrangian points (Nesvorny et al., 2013; see chapter by Emery et al.).{So far, no other model is capable to achieve these results.}  For all these reasons, nowadays simulated Solar System evolutions are required to show a jumping-Jupiter evolution to be declared successful (Nesvorny and Morbidelli, 2012). 

Although it avoids secular resonances sweeping across the asteroid belt, a jumping-Jupiter evolution is not without consequences for the asteroids. The sudden change in the eccentricity of Jupiter (from an initial basically circular orbit, {like that observed in hydrodynamical simulations of the four giant planets evolving in the gaseous proto-planetary disk,} to one with current eccentricity) changes the forced eccentricity felt by the asteroids in their secular evolution. Consequently, the proper eccentricities of the asteroids are changed. Depending on the value of the longitude of perihelion when the forced eccentricity changes, the proper eccentricity of an asteroid can decrease or increase. Roughly, 50\% of the asteroids are kicked to larger eccentricities and therefore are removed from the asteroid belt. The remaining 50\% of the asteroids have their eccentricities reduced. This can potentially reconcile the eccentricity distribution of asteroids at the end of the Grand Tack evolution (see Fig. 6) with the current distribution. Indeed, Minton and Malhotra (2011) showed that the current eccentricity distribution can be achieved starting from a primordial distribution peaked at $e\sim$0.3, {similar to that produced by the Grand Tack evolution of Jupiter (Fig. 6).}. They obtained this result using secular resonance sweeping, but the basic result should hold also for a sudden enhancement of Jupiter's eccentricity. Nevertheless specific numerical simulations have never been done to demonstrate that the eccentricity distribution of asteroids at the end of the Grand Tack model can be transformed into the current distribution via the jumping-Jupiter evolution. The jump in Jupiter's inclination due to its encounter(s) with another planet should also have partially reshuffled the asteroid inclination distribution, possibly reconciling the final distribution in Wetherill's model (see Fig. 4) with the current one. However, the effects on the inclination during Jupiter's jump seem less pronounced than those on the eccentricity (Morbidelli et al., 2010). 

The current inner edge of the asteroid belt is marked by the presence of the secular resonance between the precession frequency of the perihelion of an asteroid and the $g_6$ planetary frequency. The resonance makes unstable all objects inside 2.1 at low to moderate inclinations, which truncates the belt at its current edge. But before the impulsive separation between the orbits of Jupiter and Saturn, this resonance was much weaker (because the $g_6$  mode was less excited in the planetary system, the giant planets' orbits being more circular) and located away from the asteroid belt. Thus, in principle the asteroid belt might have extended closer to Mars before the giant planet instability event. Bottke et al. (2012) showed that the destabilization of this extended belt -or "E-belt"- could have dominated the formation of impact basins on the Moon, producing 12-15 basins over a 400~My interval. Given the age of the Orientale basin (usually estimated 3.7-3.8 Gy), this implies that the giant planet instability and the destabilization of the E-belt happened 4.1-4.2 Gy ago and was responsible for the production of the last dozen lunar basins known as the Nectarian and post-Nectarian basins.  Earlier basins and craters would have to come from other sources, such as the planetesimals leftover from the terrestrail planet formation process. The existence of two populations of projectiles, namely the left-over planetesimals dominating the bombardment at early times and the E-belt asteroids dominating the impact rate at a later epoch, should have produced a sawtooth-shaped bombardment history of the Moon (Morbidelli et al., 2012).  The E-belt should also have caused a long, slowly decaying tail of Chicxulub-sized impacts on the Earth, possibly continuing until $\sim$2~Gy ago.  Evidence for this long tail in the time-distribution of impacts is provided by the existence of terrestrial impact spherule beds, which are globally-distributed ejecta layers created by the formation of Chicxulub-sized or larger craters {(Johnson and Melosh, 2012)}: 10,  4, and 1 of these beds have been found on well-preserved, non-metamorphosed terrains between 3.23-3.47-Ga, 2.49-2.63 Ga, and 1.7-2.1 Ga, respectively (Simonson and Glass, 2004; Bottke et al., 2012).

Moreover, the escape to high-eccentricity orbits of bodies from the main belt and E-belt regions produced a spike in the impact velocities on main belt asteroids at the time of the giant planet instability. Thus, although the impact frequency on asteroids decreased with the depletion of 50\% of the main belt population and 100\% of the E-belt population, the production of impact melt on asteroids increased during this event because melt production is very sensitive to impact velocities (Marchi et al., 2013). For this reason, the impact ages of meteorites show a spike at 4.1 Gy like the lunar rocks, although for the latter this is due to a surge in the impact rate. 

A last consequence of the giant planet instability on the asteroid belt is the capture into its outer region of planetesimals from the trans-Neptunian disk (Levison et al., 2009). Because Jupiter Trojans are captured in the same event, these last captured asteroids should have had the same source as the Trojans and therefore they should be mostly of taxonomic D- and P-types. The probability of capture in the asteroid belt is nevertheless small, so it is unlikely that an object as big as Ceres was trapped from the trans-Neptunian region in this event.

\subsection{\textbf{After the giant planet instability}}
\label{afterNice}

After the giant planet instability the orbits of all planets, giants and terrestrial, are similar to the current ones (within the range of semi major axis, eccentricity and inclination oscillation provided by the current dynamical evolution). Thus, the asteroid belt has finished evolving substantially under the effect of external events such as the giant planets migration or instability. 

The asteroid belt thus entered into the current era of evolution. Asteroids became depleted at the locations of unstable resonances (mean motion and secular) on timescales that varied from resonance to resonance.  In this way, the asteroid belt acquired its current final orbital structure. In this process, it is likely that another $\sim$50\% of the asteroids were removed from the belt, most of them during the first 100~My after the giant planet instability (Minton and Malhotra, 2010). Combining this 50\% with the 50\% loss during the instability itself is the reason that we require that the primordial depletion event (Wetherill's model or Grand Tack) left a population of asteroids in the belt that was about 4x the current one. 

With the depletion of unstable resonances, the asteroid belt would have become an extraordinary boring place from the dynamical point of view. Fortunately, collisional break-up events keep refreshing the asteroid population, generating dynamical families very rich in small objects, while non-gravitational forces, mostly the Yarkovsky effect (Bottke et al., 2006) cause small asteroids to drift in semi major axis, eventually supplying new bodies to the unstable resonances. This combination of collisional activity and non-gravitational forces allow the main asteroid belt to resupply and sustain in a quasi-steady state the intrinsically unstable population of Near Earth Objects. But this is the subject of another chapter.

\section{\textbf{Conclusions and implications}}

In this chapter, we have reviewed our current understanding of the evolution of the asteroid belt, from a massive and dynamically quiet disk of planetesimals to its current state, which is so complex and rich from the points of view of both its dynamical and physical structures.  

According to this understanding, the asteroid population mainly evolved in two steps. There was an early event of strong dynamical excitation and asteroid removal, which left about 4 times the current asteroid population on orbits with a wide range of eccentricities and inclinations. This event may have been due to the self-stirring of a population of planetary embryos resident in the asteroid belt (Wetherill, 1992) or to the migration of Jupiter through the asteroid belt (the Grand Tack scenario; Walsh et al., 2011). The second step occurred later, possibly as late as 4.1 Gy ago or $\sim$400~My after the removal of gas from the proto-planetary disk. At that time, the asteroid belt underwent a second dynamical excitation and depletion when the giant planets became temporarily unstable and their orbits evolved from an initial resonant and compact configuration to the current configuration. During this second event, the asteroid belt lost about 50\% of its members. After this second event, the asteroid belt structure settled down with the progressive depletion at unstable resonances with the giant and terrestrial planets. Another 50\% of the asteroid population was lost in this process, mostly during the subsequent 100~My. 

If the first evolutionary step was due to the Grand Tack migration of Jupiter, we expect that S-type asteroids formed more or less in situ (2-3 AU region); the C-type asteroids formed in the giant planet region {(roughly 3--15 AU)} and the P- and D-type asteroids formed beyond the initial location of Neptune {(roughly 15--30 AU)}. The hot population of the Kuiper belt, the scattered disk and the Oort cloud would also derive from the same trans-Neptunian disk (Levison et al., 2008; Brasser and Morbidelli, 2013). There is a growing consensus that the cold Kuiper belt {(42--45 AU)} is primordial and born in situ (Petit et al., 2011; Parker et al., 2011;  Batygin et al., 2011; Fraser et al., 2014). Thus, the cold Kuiper belt objects should not have any correspondent in the asteroid belt. 

If instead the first step was due to the self-stirring of resident embryos as in Wetherill's model, we expect that S-type asteroids formed in the inner part of the belt, C-type asteroids in the outer part and that no asteroids sample the planetesimal population in the giant planet region. The origin of P- and D-type asteroids would be the same as above. 
Thus, deciding which of these two models is preferable requires understanding better the nature of C-type asteroids, their water content, their similarities and differences with comets and of comets among themselves.  This may not be an easy job to do. The population of main belt comets (asteroids showing cometary activity such as 133P/Elst-Pizarro or 238P/Read) and their relationship with the parent bodies of carbonaceous chondrites is key in this respect.  If it turns out that main belt comets are consistent with carbonaceous chondrites in terms of isotope composition (mostly for H, N and O) then this will argue that carbonaceous chondrites are just the rocky counterpart of bodies much richer in water/ice than meteorites themselves. This would imply that C-type asteroids formed beyond the snowline, thus presumably in the vicinity of the giant planets. If instead the main belt comets are not related to carbonaceous chondrites, but are more similar to comets from their isotope composition (it should be noticed that even though comet Hartley~II has a chondritic water D/H ratio it has a non-chondritic  $^{14}$N/$^{15}$N ratio), then this would argue for their injection in the belt from the cometary disk and suggest that the parent bodies of carbonaceous chondrites are not so water-rich and therefore they formed somewhat closer to the Sun than the snowline. 

Whatever the preferred scenario for the first depletion and excitation of the asteroid belt, it is clear that the asteroid population must have suffered in the first hundreds of My as much collisional evolution as over the last 4 Gy. However, all asteroid families formed during  the early times are not identifiable today because the dynamical excitation events dispersed them (and possibly depleted them) too severely. The presence of metallic  asteroids not associated with a family of objects of basaltic or dunitic nature, as well as the existence of rogue basaltic asteroids such as 1459 Magnya should therefore not be a surprise. The only families that are preserved are those that formed after the last giant planet instability event and that have not been made unrecognizable by subsequent collisional evolution and Yarkovsky drift; thus they are either relatively young or big. 

In this chapter, we have also examined several other asteroid excitation and depletion scenarios, most of  which have serious difficulties in reconciling their predicted outcomes with observations. We have done this not just for historical completeness, but also to illustrate the critical constraints on putative alternative scenarios of Solar System evolution. For instance, numerous studies on the possible in-situ formation of extrasolar super-Earths close to their host stars assume a large pile-up of drifting material of various sizes, from grains to small mass embryos, in the inner part of the prtoplanetary disks  (Hansen, 2014; Chatterjee and Tan; 2014; Boley et al., 2014). By analogy, these models could be used to suggest that the outer edge of the planetesimal disk at 1 AU, required to form a small Mars, was due to the same phenomenon rather than to the Grand Tack migration of Jupiter. However, from what we reported in this chapter, we think that the asteroid belt rules out this possibility. In fact, the inward migration of small planetesimals (due to gas drag) and large embryos (due to disk tides) could explain the pile-up of solid mass inside 1~AU and the mass deficit of the asteroid belt, but not the asteroids' orbital distribution. In the absence of the Grand Tack migration of Jupiter, we showed in section 3 that the only mechanism that could give to the belt an orbital structure similar to the observed one is Wetherill's model of mutual scattering of resident embryos. But if this was the case, then the mass distribution could not be concentrated within 1 AU because a massive population of embryos is required in the main belt region. Thus, at the current state of knowledge (which may change in the future), only the Grand Tack scenario seems able to explain the required mass concentration to make a small Mars.
 
In summary the asteroid belt remains the population of choice to test old, current and future models of solar system evolution.

\textbf{ Acknowledgments.} 

A.M.. was supported by the European Research Council (ERC) Advanced Grant "ACCRETE" (contract number 290568).

\bigskip

\centerline\textbf{ REFERENCES}
\bigskip
\parskip=0pt
{\small
\baselineskip=11pt

\refs Abbas, M.~M., and 11 
colleagues 2010.\ D/H ratio of Titan from Observations of the 
Cassini/Composite Infrared Spectrometer.\ The Astrophysical Journal 708, 
342-353. 

\refs  Agnor, C.~B., Lin, 
D.~N.~C.\ 2012.\ On the Migration of Jupiter and Saturn: Constraints from 
Linear Models of Secular Resonant Coupling with the Terrestrial Planets.\ 
The Astrophysical Journal 745, 143. 

\refs Alexander, 
C.~M.~O.~D.., Newsome, S.~D., Fogel, M.~L., Nittler, L.~R., Busemann, H., 
Cody, G.~D.\ 2010.\ Deuterium enrichments in chondritic macromolecular 
material{-}Implications for the origin and evolution of organics, 
water and asteroids.\ Geochimica et Cosmochimica Acta 74, 4417-4437. 

\refs Alexander, 
C.~M.~O.~D., Bowden, R., Fogel, M.~L., Howard, K.~T., Herd, C.~D.~K., 
Nittler, L.~R.\ 2012.\ The Provenances of Asteroids, and Their 
Contributions to the Volatile Inventories of the Terrestrial Planets.\ 
Science 337, 721-723. 

\refs Alexander, R., Pascucci, I., Andrews, S., Armitage, P., Cieza L., 2014. The dispersal of protoplanetary discs. In Protostars and Planets VI, in press. 

\refs Altwegg, K. and  32 co-authors, 2014. 67P/Churyumov-Gerasimenko, a Jupiter family comet with a high D/H ratio. DOI:10.1126/science.1261952

\refs Batygin, K., Brown, 
M.~E.\ 2010.\ Early Dynamical Evolution of the Solar System: Pinning Down 
the Initial Conditions of the Nice Model.\ The Astrophysical Journal 716, 
1323-1331. 

\refs Batygin, K., Brown, 
M.~E., Fraser, W.~C.\ 2011.\ Retention of a Primordial Cold Classical 
Kuiper Belt in an Instability-Driven Model of Solar System Formation.\ The 
Astrophysical Journal 738, 13. 

\refs Batygin, K., Brown, 
M.~E., Betts, H.\ 2012.\ Instability-driven Dynamical Evolution Model of a 
Primordially Five-planet Outer Solar System.\ The Astrophysical Journal 
744, L3. 

\refs Binzel, R.~P., Bus, 
S.~J., Burbine, T.~H., Sunshine, J.~M.\ 1996.\ Spectral Properties of 
Near-Earth Asteroids: Evidence for Sources of Ordinary Chondrite 
Meteorites.\ Science 273, 946-948. 

\refs Bitsch, B., Morbidelli, A., Lega, E., Crida, A.\ 2014.\ Stellar irradiated discs and implications on migration of embedded planets. II. Accreting-discs.\ Astronomy and Astrophysics 564, AA135. 

\refs Biver, N., Bockel{\'e}e-Morvan, D., Crovisier, J., Lis, D.~C., Moreno, R., Colom, P., Henry, F., Herpin, F., Paubert, G., Womack, M.\ 2006.\ Radio wavelength molecular observations of comets C/1999 T1 (McNaught-Hartley), C/2001 A2 (LINEAR), C/2000 WM$_{1}$ (LINEAR) and 153P/Ikeya-Zhang.\ Astronomy and Astrophysics 449, 1255-1270. 

\refs Bockel{\'e}e-Morvan, D., and 21 colleagues 2012.\ Herschel measurements of the D/H and $^{16}$O/$^{18}$O ratios in water in the Oort-cloud comet C/2009 P1 (Garradd).\ Astronomy and Astrophysics 544, LL15. 

\refs Boley, A.~C., Morris, 
M.~A., Ford, E.~B.\ 2014.\ Overcoming the Meter Barrier and the Formation 
of Systems with Tightly Packed Inner Planets (STIPs).\ The Astrophysical 
Journal 792, L27.

\refs Bottke, W.~F., Durda, 
D.~D., Nesvorn{\'y}, D., Jedicke, R., Morbidelli, A., Vokrouhlick{\'y}, D., 
Levison, H.\ 2005a.\ The fossilized size distribution of the main asteroid 
belt.\ Icarus 175, 111-140. 

\refs Bottke, W.~F., Durda, 
D.~D., Nesvorn{\'y}, D., Jedicke, R., Morbidelli, A., Vokrouhlick{\'y}, D., 
Levison, H.~F.\ 2005b.\ Linking the collisional history of the main asteroid 
belt to its dynamical excitation and depletion.\ Icarus 179, 63-94. 

\refs Bottke, W.~F., Jr., 
Vokrouhlick{\'y}, D., Rubincam, D.~P., Nesvorn{\'y}, D.\ 2006.\ The 
Yarkovsky and Yorp Effects: Implications for Asteroid Dynamics.\ Annual 
Review of Earth and Planetary Sciences 34, 157-191. 

\refs Bottke, W.~F., Levison, 
H.~F., Nesvorn{\'y}, D., Dones, L.\ 2007.\ Can planetesimals left over from 
terrestrial planet formation produce the lunar Late Heavy Bombardment?.\ 
Icarus 190, 203-223. 

\refs Bottke, W.~F., 
Vokrouhlick{\'y}, D., Minton, D., Nesvorn{\'y}, D., Morbidelli, A., 
Brasser, R., Simonson, B., Levison, H.~F.\ 2012.\ An Archaean heavy 
bombardment from a destabilized extension of the asteroid belt.\ Nature 
485, 78-81. 

\refs Brasser, R., Duncan, 
M.~J., Levison, H.~F.\ 2007.\ Embedded star clusters and the formation of 
the Oort cloud. II. The effect of the primordial solar nebula.\ Icarus 191, 
413-433. 

\refs Brasser, R., Duncan, 
M.~J., Levison, H.~F.\ 2008.\ Embedded star clusters and the formation of 
the Oort cloud. III. Evolution of the inner cloud during the Galactic 
phase.\ Icarus 196, 274-284. 

\refs Brasser, R., Morbidelli, A., Gomes, R., Tsiganis, K., Levison, H.~F.\ 2009.\ Constructing the secular architecture of the solar system II: the terrestrial planets.\ Astronomy and Astrophysics 507, 1053-1065. 

\refs Brasser, R., Duncan, 
M.~J., Levison, H.~F., Schwamb, M.~E., Brown, M.~E.\ 2012.\ Reassessing the 
formation of the inner Oort cloud in an embedded star cluster.\ Icarus 217, 
1-19. 

\refs Brasser, R., 
Morbidelli, A.\ 2013.\ Oort cloud and Scattered Disc formation during a 
late dynamical instability in the Solar System.\ Icarus 225, 40-49. 


\refs Burbine, T.~H., McCoy, 
T.~J., Meibom, A., Gladman, B., Keil, K.\ 2002.\ Meteoritic Parent Bodies: 
Their Number and Identification.\ Asteroids III 653-667. . 

\refs Bus, S.~J., Binzel, 
R.~P.\ 2002.\ Phase II of the Small Main-Belt Asteroid Spectroscopic 
Survey. A Feature-Based Taxonomy.\ Icarus 158, 146-177. 

\refs Busemann, H., Young, 
A.~F., O'D.~Alexander, C.~M., Hoppe, P., Mukhopadhyay, S., Nittler, L.~R.\ 
2006.\ Interstellar Chemistry Recorded in Organic Matter from Primitive 
Meteorites.\ Science 312, 727-730. 

\refs Campins, H., Hargrove, 
K., Pinilla-Alonso, N., Howell, E.~S., Kelley, M.~S., Licandro, J., 
Moth{\'e}-Diniz, T., Fern{\'a}ndez, Y., Ziffer, J.\ 2010.\ Water ice and 
organics on the surface of the asteroid 24 Themis.\ Nature 464, 1320-1321. 

\refs Carvano, J.~M., Hasselmann, P.~H., Lazzaro, D., Moth{\'e}-Diniz, T.\ 2010.\ SDSS-based taxonomic classification and orbital distribution of main belt asteroids.\ Astronomy and Astrophysics 510, A43.

\refs Chambers, J. E. and Wetherill, G. W. 1998. Making the Terrestrial Planets: N-Body Integrations of Planetary Embryos in Three Dimensions. Icarus, 136, 304-327.

\refs Chatterjee, S., 
Tan, J.~C.\ 2014.\ Inside-out Planet Formation.\ The Astrophysical Journal 
780, 53.

\refs Cohen, B.~A., Swindle, T.~D., Kring, D.~A.\ 2005.\ Geochemistry and 40Ar-39Ar geochronology of impact-melt clasts in feldspathic lunar meteorites: Implications for lunar bombardment history.\ Meteoritics and Planetary Science 40, 755. 

\refs Coustenis, A., and 10 
colleagues 2008.\ Detection of C $_{2}$HD and the D/H ratio on Titan.\ 
Icarus 197, 539-548. 

\refs Cresswell, P. and Nelson, R. P. (2008). Three-dimensional simulations of multiple protoplanets embedded in a protostellar disc. A\&A, 482, 677-690.

\refs {\'C}uk, M.\ 2012.\ Chronology 
and sources of lunar impact bombardment.\ Icarus 218, 69-79.

\refs D'Angelo, G., 
Marzari, F.\ 2012.\ Outward Migration of Jupiter and Saturn in Evolved 
Gaseous Disks.\ The Astrophysical Journal 757, 50.

\refs DeMeo, F.~E., Binzel, 
R.~P., Slivan, S.~M., Bus, S.~J.\ 2009.\ An extension of the Bus asteroid 
taxonomy into the near-infrared.\ Icarus 202, 160-180. 

\refs DeMeo, F.~E., Carry, 
B.\ 2014.\ Solar System evolution from compositional mapping of the 
asteroid belt.\ Nature 505, 629-634. 

\refs Dones, L., Weissman, 
P.~R., Levison, H.~F., Duncan, M.~J.\ 2004.\ Oort cloud formation and 
dynamics.\ Comets II 153-174. 

\refs    Duprat, J.,  Dobrica, E.,    Engrand, C.    Aleon, J.,    Marrocchi, Y.,    Mostefaoui, S.,    Meibom, A.,    Leroux, H.    Rouzaud, J. N.,    Gounelle, M.    Robert, F., 2010. Extreme Deuterium Excesses in Ultracarbonaceous Micrometeorites from Central Antarctic, Snow. Science, {\bf 1126}, 742-745.

\refs Fassett, C. I. and Minton, D. A., 2013. Impact bombardment of the terrestrial planets and the early history of the Solar System. Nature Geoscience, {\bf    6},    520-524

\refs Fischer, R.~A., Ciesla, F.~J.\ 2014.\ Dynamics of the terrestrial planets from a large number of N-body simulations.\ Earth and Planetary Science Letters 392, 28-38. 

\refs Fraser, W.~C., Brown, 
M.~E., Morbidelli, A., Parker, A., Batygin, K.\ 2014.\ The Absolute 
Magnitude Distribution of Kuiper Belt Objects.\ The Astrophysical Journal 
782, 100. 

\refs Gounelle, M., Spurn{\'y}, P., Bland, P.~A.\ 2006.\ The orbit and atmospheric trajectory of the Orgueil meteorite from historical records.\ Meteoritics and Planetary Science 41, 135-150. 

\refs Gounelle, M., 
Morbidelli, A., Bland, P.~A., Spurny, P., Young, E.~D., Sephton, M.\ 2008.\ 
Meteorites from the Outer Solar System?.\ The Solar System Beyond Neptune 
525-541. 

\refs Gradie, J., 
Tedesco, E.\ 1982.\ Compositional structure of the asteroid belt.\ Science 
216, 1405-1407. 

\refs Haisch, Jr., K. E., Lada, E. A., and Lada, C. J. (2001). Disk Frequencies and Lifetimes in Young Clusters. ApJ , 553, L153-L156.

\refs Halliday, A.~N.\ 2013.\ The 
origins of volatiles in the terrestrial planets.\ Geochimica et 
Cosmochimica Acta 105, 146-171. 

\refs Hansen, B.M.S. 2009. Formation of the Terrestrial Planets from a Narrow Annulus. The Astrophysical Journal 703, 1131-1140. 

\refs Hansen, B.~M.~S.\ 2014.\ The 
circulation of dust in protoplanetary discs and the initial conditions of 
planet formation.\ Monthly Notices of the Royal Astronomical Society 440, 
3545-3556. 

\refs Hargrove, K.~D., 
Kelley, M.~S., Campins, H., Licandro, J., Emery, J.\ 2012.\ Asteroids (65) 
Cybele, (107) Camilla and (121) Hermione: Infrared spectral diversity among 
the Cybeles.\ Icarus 221, 453-455. 

\refs Hartogh, P., and 12 
colleagues 2011.\ Ocean-like water in the Jupiter-family comet 103P/Hartley 
2.\ Nature 478, 218-220. 

\refs Heppenheimer, T. A. (1980). Secular resonances and the origin of eccentricities of Mars and the asteroids. Icarus, 41, 76-88.

\refs Hiroi, T., Zolensky, 
M.~E., Pieters, C.~M.\ 2001.\ The Tagish Lake Meteorite: A Possible Sample 
from a D-Type Asteroid.\ Science 293, 2234-2236. 

\refs Hsieh, H.~H., Jewitt, 
D.\ 2006.\ A Population of Comets in the Main Asteroid Belt.\ Science 312, 
561-563. 

\refs Ip, W.-H. (1987). Gravitational Stirring of the Asteroid Belt by Jupiter Zone Bodies. Beitraegezur Geophysik, 96, 44-51.

\refs Ishii, H.~A., Bradley, 
J.~P., Dai, Z.~R., Chi, M., Kearsley, A.~T., Burchell, M.~J., Browning, 
N.~D., Molster, F.\ 2008.\ Comparison of Comet 81P/Wild 2 Dust with 
Interplanetary Dust from Comets.\ Science 319, 447. 

\refs Jacobson, S.~A., 
Morbidelli, A., Raymond, S.~N., O'Brien, D.~P., Walsh, K.~J., Rubie, D.~C.\ 
2014.\ Highly siderophile elements in Earth's mantle as a clock for the 
Moon-forming impact.\ Nature 508, 84-87. 

\refs Jacobson, 
S.~A., Morbidelli, A.\ 2014.\ Lunar and terrestrial planet formation in the 
Grand Tack scenario.\ Royal Society of London Philosophical Transactions 
Series A 372, 174. 

\refs Jacobson, S.~A., and Walsh, K.~J. 2015. The Earth and Terrestrial Planet Formation. in press in "The Early Earth", J. Badro and M. Walter Eds.

\refs Johnson, B.~C., 
Melosh, H.~J.\ 2012.\ Impact spherules as a record of an ancient heavy 
bombardment of Earth.\ Nature 485, 75-77. 

\refs Kenyon, S. J. and Hartmann, L. (1995). Pre-Main-Sequence Evolution in the Taurus-Auriga Molecular Cloud. ApJS, 101, 11.

\refs Kley, W. and Nelson, R. P. (2012). Planet-Disk Interaction and Orbital Evolution. ARA\&A, 50, 211-249.

\refs Krasinsky, G.A., Pitjeva, E.V., Vasilyev, M.V., Yagudina, E.I. 2002. Hidden Mass in the Asteroid Belt. Icarus 158, 98-105. 

\refs Krot, S. 2014. Communication at ACM2014

\refs Kuchynka, P., 
Folkner, W.~M.\ 2013.\ A new approach to determining asteroid masses from 
planetary range measurements.\ Icarus 222, 243-253. 

\refs K{\"u}ppers, M., 
and 12 colleagues 2014.\ Localized sources of water vapour on the dwarf 
planet (1)Ceres.\ Nature 505, 525-527. 

\refs Lecar, M. and Franklin, F. (1997). The Solar Nebula, Secular Resonances, Gas Drag, and the Asteroid Belt. Icarus, 129, 134-146.

\refs Lemaitre, A. and Dubru, P. (1991). Secular resonances in the primitive solar nebula. Celestial Mechanics and Dynamical Astronomy, 52, 57-78.

\refs Levison, H.~F., 
Morbidelli, A., Van Laerhoven, C., Gomes, R., Tsiganis, K.\ 2008.\ Origin 
of the structure of the Kuiper belt during a dynamical instability in the 
orbits of Uranus and Neptune.\ Icarus 196, 258-273.

\refs Levison, H.~F., Bottke, 
W.~F., Gounelle, M., Morbidelli, A., Nesvorn{\'y}, D., Tsiganis, K.\ 2009.\ 
Contamination of the asteroid belt by primordial trans-Neptunian objects.\ 
Nature 460, 364-366. 

\refs Levison, H.~F., 
Morbidelli, A., Tsiganis, K., Nesvorn{\'y}, D., Gomes, R.\ 2011.\ Late 
Orbital Instabilities in the Outer Planets Induced by Interaction with a 
Self-gravitating Planetesimal Disk.\ The Astronomical Journal 142, 152. 

\refs Marchi, S., and 10 
colleagues 2013.\ High-velocity collisions from the lunar cataclysm 
recorded in asteroidal meteorites.\ Nature Geoscience 6, 303-307. 

\refs Marty, B.\ 2012.\ The origins and concentrations of water, carbon, nitrogen and noble gases on Earth.\ Earth and Planetary Science Letters 313, 56-66. 

\refs Masset, F., Snellgrove, M. 2001. Reversing type II migration: resonance trapping of a lighter giant protoplanet. Monthly Notices of the Royal Astronomical Society 320, L55-L59. 

\refs Meech, K.~J., and 196 
colleagues 2011.\ EPOXI: Comet 103P/Hartley 2 Observations from a Worldwide 
Campaign.\ The Astrophysical Journal 734, L1. 

\refs Minton, D.~A., 
Malhotra, R.\ 2009.\ A record of planet migration in the main asteroid 
belt.\ Nature 457, 1109-1111. 

\refs Minton, D.~A., 
Malhotra, R.\ 2010.\ Dynamical erosion of the asteroid belt and 
implications for large impacts in the inner Solar System.\ Icarus 207, 
744-757. 

\refs Minton, D.~A., 
Malhotra, R.\ 2011.\ Secular Resonance Sweeping of the Main Asteroid Belt 
During Planet Migration.\ The Astrophysical Journal 732, 53. 

\refs Morbidelli, A., Crida, A. 2007. The dynamics of Jupiter and Saturn in the gaseous protoplanetary disk. Icarus 191, 158-171. 

\refs Morbidelli, A., 
Tsiganis, K., Crida, A., Levison, H.~F., Gomes, R.\ 2007.\ Dynamics of the 
Giant Planets of the Solar System in the Gaseous Protoplanetary Disk and 
Their Relationship to the Current Orbital Architecture.\ The Astronomical 
Journal 134, 1790-1798. 

\refs Morbidelli, A., 
Bottke, W.~F., Nesvorn{\'y}, D., Levison, H.~F.\ 2009.\ Asteroids were born 
big.\ Icarus 204, 558-573. 

\refs Morbidelli, A., 
Brasser, R., Gomes, R., Levison, H.~F., Tsiganis, K.\ 2010.\ Evidence from 
the Asteroid Belt for a Violent Past Evolution of Jupiter's Orbit.\ The 
Astronomical Journal 140, 1391-1401. 

\refs Morbidelli, A., Marchi, S., Bottke, W.~F., Kring, D.~A.\ 2012.\ A sawtooth-like timeline for the first billion years of lunar bombardment.\ Earth and Planetary Science Letters 355, 144-151. 

\refs Moth{\'e}-Diniz, T., Carvano, J.~M.~{\'a}., Lazzaro, D.\ 2003.\ 
Distribution of taxonomic classes in the main belt of asteroids.\ Icarus 
162, 10-21. 

\refs Mouillet, D., Larwood, 
J.~D., Papaloizou, J.~C.~B., Lagrange, A.~M.\ 1997.\ A planet on an 
inclined orbit as an explanation of the warp in the Beta Pictoris disc.\ 
Monthly Notices of the Royal Astronomical Society 292, 896. 

\refs Nagasawa, M., Tanaka, H., and Ida, S. (2000). Orbital Evolution of Asteroids during Depletion of the Solar Nebula. AJ , 119, 1480-1497.

\refs Nagasawa, M., Ida, S., and Tanaka, H. (2001). Origin of high orbital eccentricity and inclination of asteroids. Earth, Planets, and Space, 53, 1085-1091.

\refs Nagasawa, M., Ida, S., and Tanaka, H. (2002). Excitation of Orbital Inclinations of Asteroids during Depletion of a Protoplanetary Disk: Dependence on the Disk Configuration. Icarus, 159, 322-327.

\refs Nakamura, T., and 11 
colleagues 2008.\ Chondrulelike Objects in Short-Period Comet 81P/Wild 2.\ 
Science 321, 1664. 

\refs Nesvorn{\'y}, D., 
Vokrouhlick{\'y}, D., Morbidelli, A.\ 2007.\ Capture of Irregular 
Satellites during Planetary Encounters.\ The Astronomical Journal 133, 
1962-1976. 

\refs Nesvorn{\'y}, D., 
Jenniskens, P., Levison, H.~F., Bottke, W.~F., Vokrouhlick{\'y}, D., 
Gounelle, M.\ 2010.\ Cometary Origin of the Zodiacal Cloud and Carbonaceous 
Micrometeorites. Implications for Hot Debris Disks.\ The Astrophysical 
Journal 713, 816-836. 

\refs Nesvorn{\'y}, D.\ 2011.\ 
Young Solar System's Fifth Giant Planet?.\ The Astrophysical Journal 742, 
L22. 

\refs Nesvorn{\'y}, D., Morbidelli, A.\ 2012.\ Statistical Study of the Early 
Solar System's Instability with Four, Five, and Six Giant Planets.\ The 
Astronomical Journal 144, 117. 

\refs Nesvorn{\'y}, D., 
Vokrouhlick{\'y}, D., Morbidelli, A.\ 2013.\ Capture of Trojans by Jumping 
Jupiter.\ The Astrophysical Journal 768, 45. 

\refs Nesvorn{\'y}, D., 
Vokrouhlick{\'y}, D., Deienno, R.\ 2014.\ Capture of Irregular Satellites 
at Jupiter.\ The Astrophysical Journal 784, 22. 

\refs Niemann, H.~B., Atreya, 
S.~K., Demick, J.~E., Gautier, D., Haberman, J.~A., Harpold, D.~N., 
Kasprzak, W.~T., Lunine, J.~I., Owen, T.~C., Raulin, F.\ 2010.\ Composition 
of Titan's lower atmosphere and simple surface volatiles as measured by the 
Cassini-Huygens probe gas chromatograph mass spectrometer experiment.\ 
Journal of Geophysical Research (Planets) 115, E12006.

\refs Nixon, C.~A., and 12 
colleagues 2012.\ Isotopic Ratios in Titan's Methane: Measurements and 
Modeling.\ The Astrophysical Journal 749, 159. 

\refs O'Brien, D. P., Morbidelli, A., and Levison, H. F. (2006). Terrestrial planet formation with strong dynamical friction. Icarus, 184, 39-58.

\refs O'Brien, D. P., Morbidelli, A., and Bottke, W. F. (2007). The primordial excitation and clearing of the asteroid belt-Revisited. Icarus, 191, 434-452.

\refs Parker, A.~H., 
Kavelaars, J.~J., Petit, J.-M., Jones, L., Gladman, B., Parker, J.\ 2011.\ 
Characterization of Seven Ultra-wide Trans-Neptunian Binaries.\ The 
Astrophysical Journal 743, 1. 

\refs Petit, J., Morbidelli, A., and Valsecchi, G. B. (1999). Large Scattered Planetesimals and the Excitation of the Small Body Belts. Icarus, 141, 367-387.

\refs Petit, J., Morbidelli, A., and Chambers, J. (2001). The Primordial Excitation and Clearing of the Asteroid Belt. Icarus, 153, 338-347.

\refs Petit, J., Chambers, J., Franklin, F., and Nagasawa, M. (2002). Primordial Excitation and Depletion of the Main Belt. In W. F. Bottke, A. Cellino, P. Paolicchi, and R. P. Binzel, editors, Asteroids III , pages 711-738. University of Arizona Press, Tucson, AZ.

\refs Petit, J.-M., and 16 
colleagues 2011.\ The Canada-France Ecliptic Plane Survey{-}Full Data 
Release: The Orbital Structure of the Kuiper Belt.\ The Astronomical 
Journal 142, 131. 

\refs Pierens, A., Nelson, R.P. 2008. Constraints on resonant-trapping for two planets embedded in a protoplanetary disc. Astronomy and Astrophysics 482, 333-340. 

\refs Pierens, A., Raymond, S.~N.\ 2011.\ Two phase, inward-then-outward migration of Jupiter and Saturn in the gaseous solar nebula.\ Astronomy and Astrophysics 533, A131. 

\refs Rivkin, A.~S., Emery, 
J.~P.\ 2010.\ Detection of ice and organics on an asteroidal surface.\ 
Nature 464, 1322-1323. 

\refs Robuchon, G., Nimmo, 
F., Roberts, J., Kirchoff, M.\ 2011.\ Impact basin relaxation at Iapetus.\ 
Icarus 214, 82-90. 

\refs Rousselot, P., and 11 
colleagues 2014.\ Toward a Unique Nitrogen Isotopic Ratio in Cometary 
Ices.\ The Astrophysical Journal 780, LL17. 


\refs Simonson, B.~M., 
Glass, B.~B.\ 2004.\ Spherule Layersrecords of Ancient Impacts.\ Annual 
Review of Earth and Planetary Sciences 32, 329-361. 

\refs Somenzi, L., Fienga, A., Laskar, J., Kuchynka, P.\ 2010.\ Determination of asteroid masses from their close encounters with Mars.\ Planetary and Space Science 58, 858-863. 

\refs Strom, S. E., Edwards, S., and Skrutskie, M. F. (1993). Evolutionary time scales for circumstellar disks associated with intermediate- and solar-type stars. In E. H. Levy \& J. I. Lunine, editor, Protostars and Planets III , pages 837-866.

\refs Tholen, D.~J.\ 1984.\ Asteroid 
taxonomy from cluster analysis of Photometry.\ Ph.D.~Thesis. 

\refs Thommes, E.~W., Duncan, 
M.~J., Levison, H.~F.\ 1999.\ The formation of Uranus and Neptune in the 
Jupiter-Saturn region of the Solar System.\ Nature 402, 635-638. 

\refs Tsiganis, K., Gomes, R., Morbidelli, A., and Levison, H. F. (2005). Origin of the orbital architecture of the giant planets of the Solar System. Nature, 435, 459-461.

\refs Villeneuve, J., 
Chaussidon, M., Libourel, G.\ 2009.\ Homogeneous Distribution of $^{26}$Al 
in the Solar System from the Mg Isotopic Composition of Chondrules.\ 
Science 325, 985. 

\refs Yang, L., Ciesla, F.~J., 
Alexander, C.~M.~O.~'.\ 2013.\ The D/H ratio of water in the solar nebula 
during its formation and evolution.\ Icarus 226, 256-267. 

\refs Waite, J.~H., Jr., and 15 
colleagues 2009.\ Liquid water on Enceladus from observations of ammonia 
and $^{40}$Ar in the plume.\ Nature 460, 487-490. 

\refs Walsh, K. J., Morbidelli, A., Raymond, S. N., O'Brien, D. P., and Mandell, A. M. (2011). A low mass for Mars from Jupiter's early gas-driven migration. Nature, 475, 206-209.

\refs Ward, W.~R., Colombo, G., 
Franklin, F.~A.\ 1976.\ Secular resonance, solar spin down, and the orbit 
of Mercury.\ Icarus 28, 441-452. 

\refs Ward, W. R. (1981). Solar nebula dispersal and the stability of the planetary system. I - Scanning secular resonance theory. Icarus, 47, 234-264.

\refs Weidenschilling, S.~J.\ 1977.\ The distribution of mass in the planetary system and solar nebula.\ Astrophysics and Space Science 51, 153-158. 

\refs Wetherill, G. W. (1992). An alternative model for the formation of the asteroids. Icarus, 100, 307-325.

\refs Zolensky, M.~E., and 
74 colleagues 2006.\ Mineralogy and Petrology of Comet 81P/Wild 2 Nucleus 
Samples.\ Science 314, 1735. 

\refs Zolensky, M., and 30 colleagues 2008.\ Comparing Wild 2 particles to chondrites and IDPs.\ Meteoritics and Planetary Science 43, 261-272. 

\refs Zuckerman, B., Forveille, T., and Kastner, J. H. (1995). Inhibition of giant-planet formation by rapid gas depletion around young stars. Nature, 373, 494-496.

\end{document}